\documentclass[prc,twocolumn,floatfix,groupedaddress,nofootinbib,showpacs,preprintnumbers,amsmath,amssymb,amsfonts,superscriptaddress,widetable,href] {revtex4}
\usepackage{footnote}
\usepackage{wasysym}
\usepackage{color}
\usepackage{graphicx}
\usepackage{dcolumn}
\usepackage{multirow}
\usepackage{bm}
\usepackage{sidecap}
\usepackage{ulem}
\usepackage{hyperref}
\usepackage{natbib}
\usepackage{url}
\usepackage{textcase}
\usepackage{bm}
\begin{document}
\title{Description of Nuclei with Magic Number Z(N) = 6}

\author{M. Kumawat}
\affiliation{Department of Physics, Govt. Women Engineering College,
Ajmer - 305002, India}
\affiliation{Department of Physics, School of Basic Sciences, Manipal
Univ., Jaipur-303007, India}
\author{G. Saxena}
\email{gauravphy@gmail.com}
\affiliation{Department of Physics, Govt. Women Engineering College,
Ajmer - 305002, India}
\affiliation{Department of Physics, School of Basic Sciences, Manipal
Univ., Jaipur-303007, India}
\author{M. Kaushik}
\affiliation{Department of Physics, Shankara Institute of Technology, Kukas, Jaipur-302028, India}
\author{R. Sharma}
\affiliation{Department of Physics, School of Basic Sciences, Manipal Univ., Jaipur-303007, India}
\author{S. K. Jain}
\affiliation{Department of Physics, School of Basic Sciences, Manipal Univ., Jaipur-303007, India}
\begin{abstract}

Encouraged with the evidence for Z = 6 'magic number' in neutron-rich carbon isotopes, we have performed relativistic mean-field plus BCS calculations to investigate ground state properties of entire chains of isotopes(isotones) with Z(N) = 6 including even and odd mass nuclei.
Our calculations include deformation, binding energy, separation energy, single particle energy, rms radii along with
charge and neutron density profile etc., and are found in an excellent match with latest experimental results demonstrating Z = 6 as a strong magic number. N = 6 is also found to own similar kind of strong magic character.

\end{abstract}
\maketitle
\section{Introduction}
There have been several experimental evidences since last two decades which have proven the evolution of nuclear shell structures and consequently visualized new magic numbers along with disappearance of conventionally known magic numbers in the various regions of periodic chart. Few of them have confirmed magicity for N = 14 \cite{stanoiu,brown,becheva}, N = 16 \cite{kanungo,hoffman,tshoo}, Z = 16 \cite{togano1}, N = 32 \cite{gade,wienholtz,rosenbusch}, N = 34 \cite{stepp} etc., and break down of conventional magicity N = 8, 20, 28 etc. \cite{iwasaki,door,bastin}.  Moreover, from separation energy systematics, beta decay Q-values and first excited states of nuclei the possibility of a new sub-shell closure at N = 6 is suggested for neutron-rich isotopes by Kanungo {\textit{et al.}} \cite{kanungo}. Furthermore, magicity of N = 6 has been discussed in terms of spin-isospin dependent part of the nucleon-nucleon interaction by Otsuka {\textit{et al.}} \cite{otsuka}. In addition, it has been shown using extension of the Bethe-Weizsäcker mass formula \cite{samanta}, Relativistic Mean-Field theory \cite{jha} and potential energy surfaces calculated within the cluster-core model \cite{gupta} that N = 6 and Z = 6 indeed show characteristics similar to shell closures. In 2008, ground state structure of $^{9}$Li has been investigated through the d(Li$^9$,t)Li$^8$ one-neutron transfer reaction at E/A = 1.68 MeV and spectroscopic factors are derived from a DWBA analysis by Kanungo {\textit{et al.}}, which showed dominance of N = 6 sub-shell gap over conventional shell gap at N = 8 \cite{kanungo1}. Y. Kanada {\textit{et al.}} has investigated neutron number dependence of proton radii of Be, B, and C isotopes and found that proton radii in C isotopes show a weak N dependence because of the stable proton structure in nuclei with Z = 6 \cite{kanada}. Recently, in 2017, experimental evidences for proton number Z = 6 in $^{13-20}$C are obtained from systematic analysis of radii, electromagnetic transition rates and atomic masses of light nuclei \cite{tran}, which have invoked this present theoretical study.

In this communication, we describe ground state properties of even and odd mass nuclei with Z = 6 and N = 6 to investigate magic character of Z(N) = 6 from neutron drip-line to proton drip-line. To probe magicity with the help of various properties viz. binding energies, single particle energies, deformations, separation energies, pairing energies as well as radii and density distributions etc., we use Relativistic Mean-Field plus Bardeen-Cooper-Schrieffer (RMF+BCS) approach \cite{walecka,boguta,bouy,pgr2,suga,ring,sharma,lala,mizu,estal,yadav,yadav1} with NL3* parameter \cite{nl3star}. The results are compared with the available experimental data \cite{wang} and results of another popular force parameter PK1~\cite{pk1} at places to testify our outcomes. In our approach single particle continuum corresponding to the Relativistic Mean-Field (RMF) is replaced by a set of discrete positive energy states \cite{yadav,yadav1} yielding results in excellent agreement with the experimental data along with recent continuum Relativistic Hartree-Bogoliubov (RCHB) and other similar mean-field calculations \cite{meng5,meng2}. This approach has proven to be very successful for the extensive study of (i) conventional and new magic nuclei \cite{yadav1,saxena,saxena1,saxena2,saxena3,saxena5}, (ii) two proton radioactivity \cite{singh}, and recently describing (iii) interdependence of 2p-halo with 2p-radioactivity \cite{saxena4} and (iv) Bubble structure \cite{saxena5}.

\section{Relativistic Mean-Field Theory}

RMF calculations have been carried out using the model Lagrangian density with nonlinear terms both for the ${\sigma}$ and ${\omega}$ mesons as described in detail in Refs.~\cite{suga,yadav,singh}.

\begin{eqnarray}
       {\cal L}& = &{\bar\psi} [\imath \gamma^{\mu}\partial_{\mu}
                  - M]\psi\nonumber\\
                  &&+ \frac{1}{2}\, \partial_{\mu}\sigma\partial^{\mu}\sigma
                - \frac{1}{2}m_{\sigma}^{2}\sigma^2- \frac{1}{3}g_{2}\sigma
                 ^{3} - \frac{1}{4}g_{3}\sigma^{4} -g_{\sigma}
                {\bar\psi}  \sigma  \psi\nonumber\\
               &&-\frac{1}{4}H_{\mu \nu}H^{\mu \nu} + \frac{1}{2}m_{\omega}
                  ^{2}\omega_{\mu}\omega^{\mu} + \frac{1}{4} c_{3}
                 (\omega_{\mu} \omega^{\mu})^{2}
                  - g_{\omega}{\bar\psi} \gamma^{\mu}\psi
                 \omega_{\mu}\nonumber\\
              &&-\frac{1}{4}G_{\mu \nu}^{a}G^{a\mu \nu}
                 + \frac{1}{2}m_{\rho}
                 ^{2}\rho_{\mu}^{a}\rho^{a\mu}
                  - g_{\rho}{\bar\psi} \gamma_{\mu}\tau^{a}\psi
                 \rho^{\mu a}\nonumber\nonumber\\
               &&-\frac{1}{4}F_{\mu \nu}F^{\mu \nu}
                 - e{\bar\psi} \gamma_{\mu} \frac{(1-\tau_{3})}
                 {2} A^{\mu} \psi\,\,,%\nonumber\
\end{eqnarray}
where the field tensors $H$, $G$ and $F$ for the vector fields are
defined by
\begin{eqnarray}
                 H_{\mu \nu} &=& \partial_{\mu} \omega_{\nu} -
                       \partial_{\nu} \omega_{\mu}\nonumber\\
                 G_{\mu \nu}^{a} &=& \partial_{\mu} \rho_{\nu}^{a} -In t
                       \partial_{\nu} \rho_{\mu}^{a}
                     -2 g_{\rho}\,\epsilon^{abc} \rho_{\mu}^{b}
                    \rho_{\nu}^{c} \nonumber\\
                  F_{\mu \nu} &=& \partial_{\mu} A_{\nu} -
                       \partial_{\nu} A_{\mu}\,\,\nonumber\
\end{eqnarray}

and other symbols have their usual meaning. Based on the
single-particle spectrum calculated by the Relativistic Mean-Field (RMF) described above, we
perform a state dependent BCS calculations \cite{lane,ring2}. The continuum is replaced by a set of positive
energy states generated by enclosing the nucleus in a spherical box. Thus the gap equations have the standard form for all the single
particle states, i.e.
\begin{eqnarray}
     \Delta_{j_1}=-\frac{1}{2}\frac{1}{\sqrt{2j_1+1}}
     \sum_{j_2}\frac{\left<{({j_1}^2)\,0^+\,|V|\,({j_2}^2)\,0^+}\right>}
      {\sqrt{\big(\varepsilon_{j_2}\,-\,\lambda \big)
       ^2\,+\,{\Delta_{j_2}^2}}}\sqrt{2j_2+1}\Delta_{j_2}\,\,
\end{eqnarray}\\
where $\varepsilon_{j_2}$ are the single particle energies, and
$\lambda$ is the Fermi energy, whereas the particle number condition
is given by $\sum_j \, (2j+1) v^2_{j}\, = \,{\rm N}$. In the
calculations, we use a zero-range delta force for the pairing interaction.
\begin{eqnarray}
V = -V_0 \delta(r)
\end{eqnarray}\\
Apart from its simplicity, the applicability and justification of using
such a $\delta$-function form of interaction has been discussed in Ref. \cite{dobac-delta}, whereby it has been shown in the
context of Hartree-Fock-Bogoliubov (HFB) calculations that the use of a delta force in a
finite space simulates the effect of finite range interaction in a
phenomenological manner (see also \cite{meng5}
for more details). In this work the pairing strength
is taken to be the same for both protons and neutrons i.e. V$_0$ = 350 MeV fm$^3$ which is used in Refs.$~$ \cite{yadav,saxena,saxena1}
for the successful description of drip-line nuclei and obtained by fitting the experimental
odd-even staggerings with the cut off $E - \lambda$ = 8.0 MeV where E is the single-particle energy and $\lambda$
the chemical potential. The pairing matrix element for the
$\delta$-function force is given by
\begin{eqnarray}
\left<{({j_1}^2)\,0^+\,|V|\,({j_2}^2)\,0^+}\right>
=-\,\frac{V_0}{8\pi}
       \sqrt{(2j_1+1)(2j_2+1)}\,\,I_R\,\,
\end{eqnarray}
where $I_R$ is the radial integral having the form
\begin{eqnarray}
   I_R& = &\,\int\,dr \frac{1}{r^2}\,\left(G^\star_{j_ 1}\, G_{j_2}\,+\,
     F^\star_{j_ 1}\, F_{j_2}\right)^2
\end{eqnarray}
Here $G_{\alpha}$ and $F_{\alpha}$ denote the radial wave functions
for the upper and lower components, respectively, of the nucleon
wave function. The coupled field equations obtained from the Lagrangian density in (1)
are finally reduced to a set of simple radial equations which are solved self consistently along with the equations for the
state dependent pairing gap $\Delta_{j}$ and the total particle
number N for a given nucleus.

The Relativistic Mean-Field description was extended to treat deformed nuclei of axially symmetric shapes by Gambhir {\textit{et al.}}
\cite{gambhir} using an expansion method which has been commonly used in various Refs.~ \cite{saxena,saxena3,singh,geng1}.
The scalar, vector, isovector and charge densities, as in the
spherical case, are expressed in terms of the spinor $\pi_i$, its
conjugate $\pi_i^+$, operator $\tau_3$ etc. These densities serve as
sources for the fields $\phi$ = $\sigma$, $\omega^0$, $\rho^0$ and
$A^0$, which are determined by the Klein-Gordon equation in
cylindrical coordinates. Thus a set of coupled equations, namely the
Dirac equation with potential terms for the nucleons and
Klein-Gordon type equations with sources for the mesons and the
photon, is obtained. These equations are solved self consistently.
For this purpose, as described above, the well-tested basis
expansion method has been employed~\cite{gambhir,geng1}.
For further details of these formulations we refer the
reader to Refs.~\cite{gambhir,singh,geng1}.

\section{Results and discussions}

\begin{figure}[h]
\centering
\includegraphics[width=0.5\textwidth]{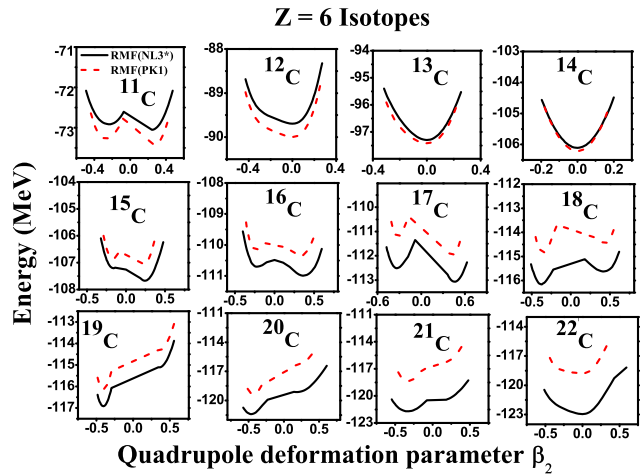}
\caption{(Colour online) The potential energy surfaces of C isotopes as a function of the deformation parameter $\beta_{2}$ calculated by NL3* and PK1 parameters.} \label{fig1} \end{figure}

As mentioned above, we have performed Relativistic Mean-Field (RMF) calculations with axially symmetric deformed shapes
\cite{gambhir,ring,singh} using NL3* \cite{nl3star} and PK1 \cite{pk1} force parameters for C isotopes to analyze magicity from ground state properties. In Fig. \ref{fig1}, we have plotted potential energy curves of $^{11-22}$C with respect to quadrupole deformation parameter $\beta_{2}$ to examine shapes in ground state. Energy curves of $^{12,13,14}$C and $^{22}$C show only one minima at $\beta_{2}$ = 0 indicating spherical shapes whereas rest isotopes are found deformed in which $^{11,15,16,17,18}$C are ascertained to exhibit shape co-existence (prolate and oblate both shapes) and $^{19,20,21}$C are found with dominant oblate shape. These observation are found consistent with earlier calculations done by NL-SH parameter by Sharma {\textit{et al.}} \cite{mmsharma} and the deformed
Hartree-Fock (HF)+ BCS model with Skyrme interaction done by Suzuki {\textit{et al.}} \cite{suzuki}. It is also noteworthy here that our results with NL3* and PK1 force parameters are in good match as both of these parameters show same kind of shapes and deformation for all C isotopes mentioned in Fig. \ref{fig1}.

\begin{figure}[h]
\centering
\includegraphics[width=0.5\textwidth]{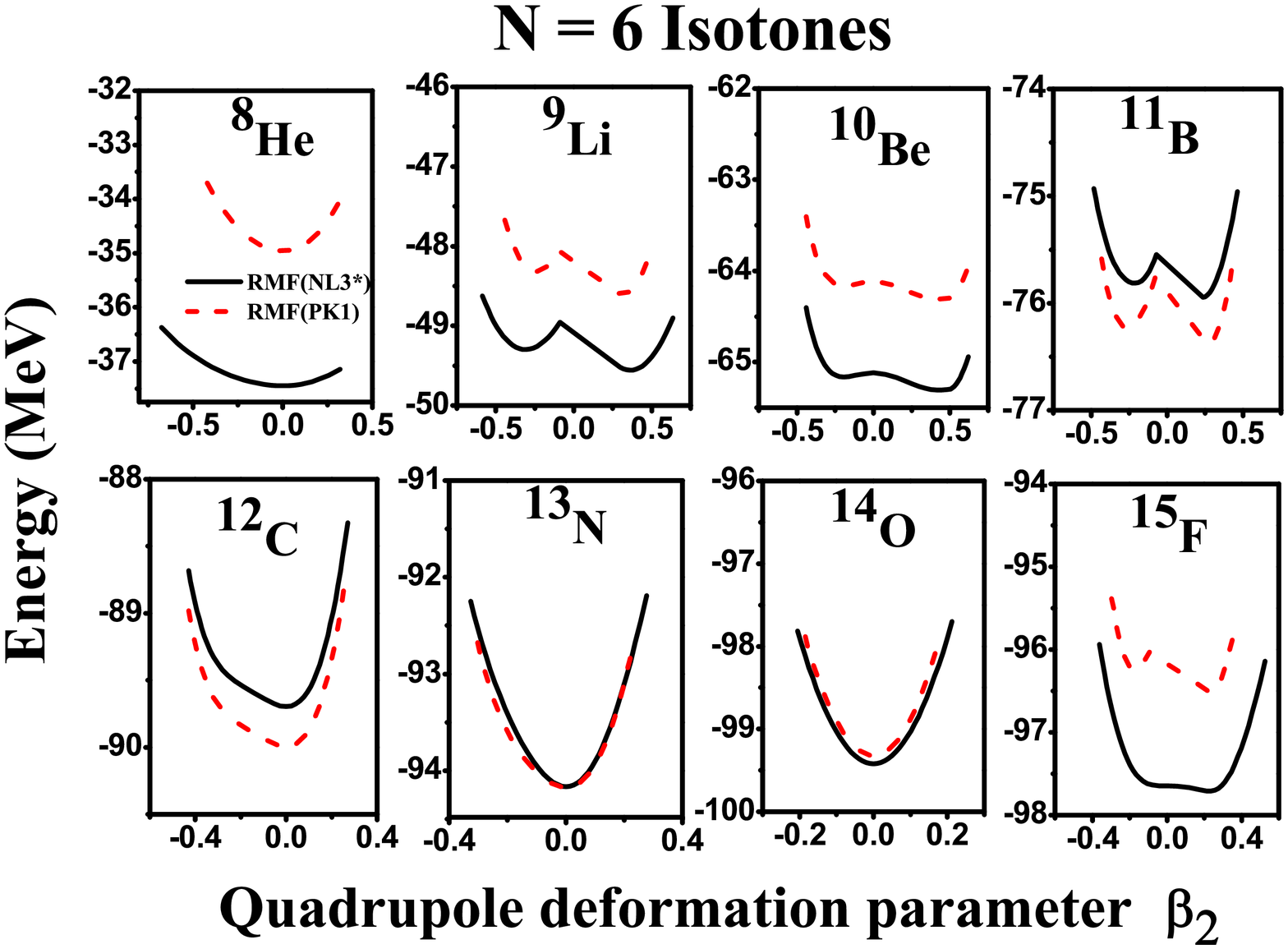}
\caption{(Colour online) Same as Fig. 1 but for N = 6 isotones.} \label{fig2} \end{figure}

In a similar manner, we have plotted potential energy curves as a function of deformation for N = 6 isotones in Fig. \ref{fig2}. Here also, we have found some spherical candidates viz. $^8$He, $^{12}$C, $^{13}$N, $^{14}$O and all others are found with prolate and oblate shapes co-existence. Interestingly, among these spherical candidates, $^8$He \cite{skaza} and $^{14}$O \cite{brown,brown1} are already reported as doubly magic nuclei. In Table I, we have tabulated quadrupole deformation obtained by our calculations and by some other theoretical predictions \cite{mmsharma,sagawa,Lalazissis}  along with energy difference between oblate and prolate minima for the nuclei showing shape coexistence from Figs. \ref{fig1} and \ref{fig2}.

\begin{table}
\caption{Quadrupole deformation $\beta_2$ and energy difference($\bigtriangleup$E) between prolate and oblate minima for the nuclei with Z(N) = 6 are shown. Other theoretical data of $\beta_2$ are taken from Refs.$~$\cite{mmsharma,sagawa,Lalazissis}.}
\centering
\resizebox{0.4\textwidth}{!}{%
{\begin{tabular}{c|c|c|c|c|c}
 \hline
 \multicolumn{1}{c|}{Nuclei}&
 \multicolumn{3}{c|}{$\beta_{2}$}&
 %\multicolumn{1}{c}{Other theories}&
 \multicolumn{2}{c}{$\bigtriangleup$E(MeV)}\\
 \hline
 \multicolumn{1}{c|}{}&
 \multicolumn{2}{c|}{RMF}&
 \multicolumn{1}{c|}{other}&
   \multicolumn{2}{c}{RMF}\\
   \cline{2-3}
 \cline{5-6}

 \multicolumn{1}{c|}{}&
 \multicolumn{1}{c|}{NL3*}&
 \multicolumn{1}{c|}{PK1}&
   \multicolumn{1}{c|}{theories}&
    \multicolumn{1}{c|}{NL3*}&
     \multicolumn{1}{c}{PK1}\\
   \hline
  $^{11}${C}& 0.243& 0.274&&0.138&0.170\\
  $^{12}${C}& 0.000& 0.000&-0.230&  -   & -    \\
  $^{13}${C}& 0.000& 0.000&&  -   & -    \\
  $^{14}${C}& 0.000& 0.000&0.000&  -   & -    \\
  $^{15}${C}& 0.246& 0.238&&0.468&0.220\\
  $^{16}${C}& 0.358& 0.336&-0.005&0.294&0.212\\
  $^{17}${C}& 0.456& 0.418&0.400&0.558&0.761\\
  $^{18}${C}&-0.380&-0.343&-0.316&0.536&0.385\\
  $^{19}${C}&-0.439&-0.403&&1.785    &1.806\\
  $^{20}${C}&-0.458&-0.435&-0.405&2.267&-    \\
  $^{21}${C}&-0.339&-0.332&& -   &-    \\
  $^{22}${C}& 0.000& 0.000&-0.308&  -   & -    \\
  \hline
  $^{8}${He}& 0.000&0.000 &&-& -        \\
  $^{9}${Li}& 0.359&0.308 &&0.260&0.281\\
  $^{10}${Be}&0.117&0.412 &&0.183&0.145\\
  $^{11}${B}& 0.117&0.281 &&0.132&0.161\\
  $^{12}${C}& 0.000&0.000 &-0.23&-&  - \\
  $^{13}${N}& 0.000&0.000 &&-&  -       \\
  $^{14}${O}& 0.000&0.000 &&-&  -       \\
  $^{15}${F}& 0.211&0.234 &&-&0.272     \\
  $^{16}${Ne}&0.444&0.345 &0.016&0.384&0.166\\
\hline
\end{tabular}}}
\end{table}

\begin{figure}[h]
\centering
\includegraphics[width=0.5\textwidth]{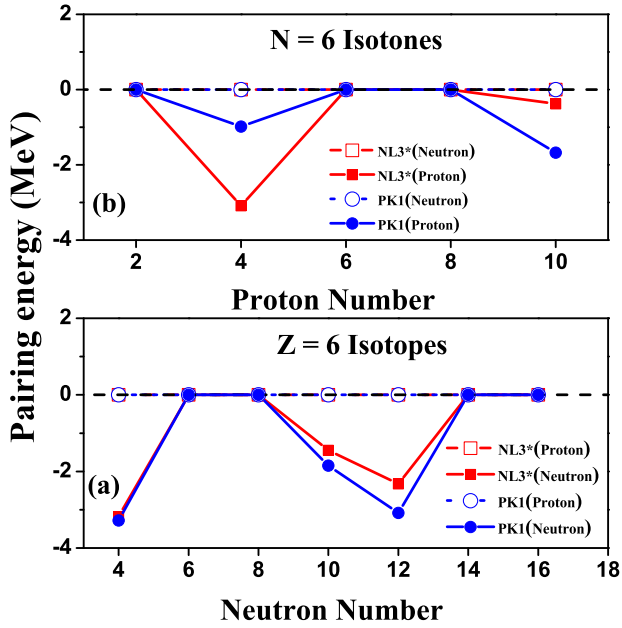}
\caption{(Colour online) Pairing energy contributions are shown separately for protons and neutrons for (a) Z= 6 isotopes and (b) N = 6 isotones calculated by NL3* and PK1 parameters.} \label{fig3} \end{figure}

Further to explore magic characteristics in these isotopes(isotones), we have depicted pairing energy contribution of protons and neutrons in Fig. \ref{fig3}, which is an indicator of shell closure if it vanishes to zero. In view of this, it is gratifying to note that proton pairing energy indeed remains zero for all the C isotopes and neutron pairing energy shows zero value for all isotones of N = 6, indicating proton and neutron magic character for full chain of nuclei, respectively. From Fig. \ref{fig3}(a), it is evident that neutron pairing energy also vanishes for N = 6, 8 and 14 refereing $^{12,14,22}$C as doubly magic nuclei. Their zero deformations as depicted in Fig. \ref{fig1} and Table I establish these nuclei as potential candidates of doubly magicity. In a similar manner from Fig. \ref{fig3}(b) and Fig. \ref{fig2} along with Table 1, we find doubly magic nuclei $^8$He, $^{12}$C and $^{14}$O. These identified doubly magic nuclei using pairing energy are in agreement with candidates reported in Refs. $~$ \cite{samanta,jha,gupta,saxena5,skaza,brown,brown1}.

\begin{figure}[h]
\centering
\includegraphics[width=0.5\textwidth]{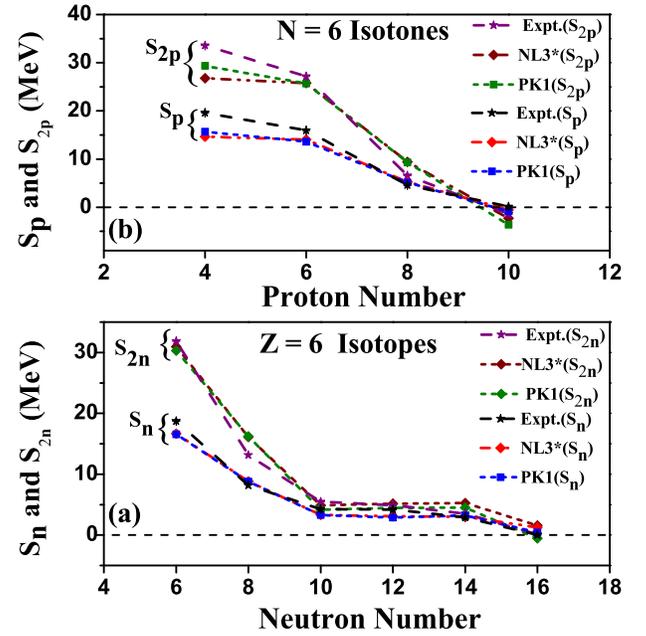}
\caption{(Colour online) Neutron separation energies (S$_{2n}$ and S$_{n}$) and proton separation energies (S$_{2p}$ and S$_{p}$) calculated with NL3* and PK1 parameters are shown for Z = 6 isotopes and N = 6 isotones along with experimental separation energies \cite{wang}.} \label{fig4} \end{figure}

For a comparison with experimental data and examining credibility of our results, in Fig. \ref{fig4}, we show neutron separation energies for C isotopes and proton separation energies for N = 6 isotones along with experimental separation energies \cite{wang}. It is indeed noteworthy here from Fig. \ref{fig4} that our calculations with both the parameters show excellent match with experimental data \cite{wang}. The shell closures as revealed by the pairing energy graph (Fig. \ref{fig3}) are also exhibited as an abrupt change in the values of two- and one-neutron (two- and one-proton) separation energies for the isotopes(isotones) next to magic numbers. From Fig. \ref{fig4}(a) an abrupt change can be seen just after N = 6, 8 and also for 16 (if plot further) \cite{saxena5} showing magic character in $^{12,14,22}$C and similarly after Z = 6 and 8 reflecting double magicity in $^{12}$C and $^{14}$O.

\begin{figure}[h]
\centering
\includegraphics[width=0.5\textwidth]{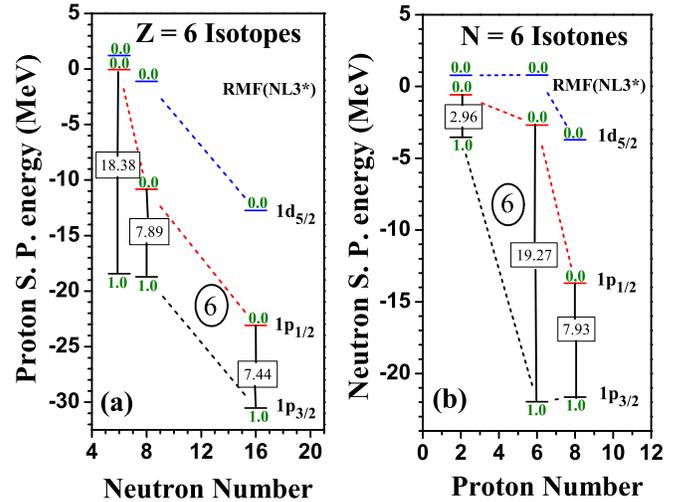}
\caption{(Colour online) Proton and neutron single particle energies for Z = 6 and N = 6 are shown in (a) and (b) respectively.} \label{fig5} \end{figure}

For further investigation of these magicity, we have calculated single particle energies using spherical framework of RMF+BCS approach \cite{saxena1,saxena2} for the nuclei which are found doubly magic and spherical from above study. With the use of spherical framework, role of single particle states and density distribution can be manifested in a much simpler and effective way \cite{saxena1,saxena2} therefore, in Figs. \ref{fig5}(a) and \ref{fig5}(b), we have shown proton and neutron single particle energies for Z = 6 and N = 6, respectively, along with occupancy of each state shown by numbers. One can find a large energy gap between proton 1p$_{3/2}$ and 1p$_{1/2}$ in Fig. \ref{fig5}(a) for all the nuclei considered here from neutron drip-line to proton drip-line. This gap which gives rise to shell closure at Z = 6, is found always $>$ 7 MeV and also which is comparable to the gap between 1p$_{1/2}$ and 1d$_{5/2}$ responsible for magic number Z = 8. Hence, these proton single particle levels demonstrate magic character of Z = 6 in accordance with latest observation \cite{tran}. In a similar manner, neutron single particle levels of N = 6 isotones provide plenty grounds of magicity of N = 6 as can be seen from Fig. \ref{fig5}(b). For full chain of isotopes(isotones) of Z(N) = 6 the occupancy of proton(neutron) single particle 1p$_{1/2}$ state is always found zero, as can be seen by numbers on the states, bespeaking shell closure for Z(N) = 6.

\begin{figure}[h]
\centering
\includegraphics[width=0.5\textwidth]{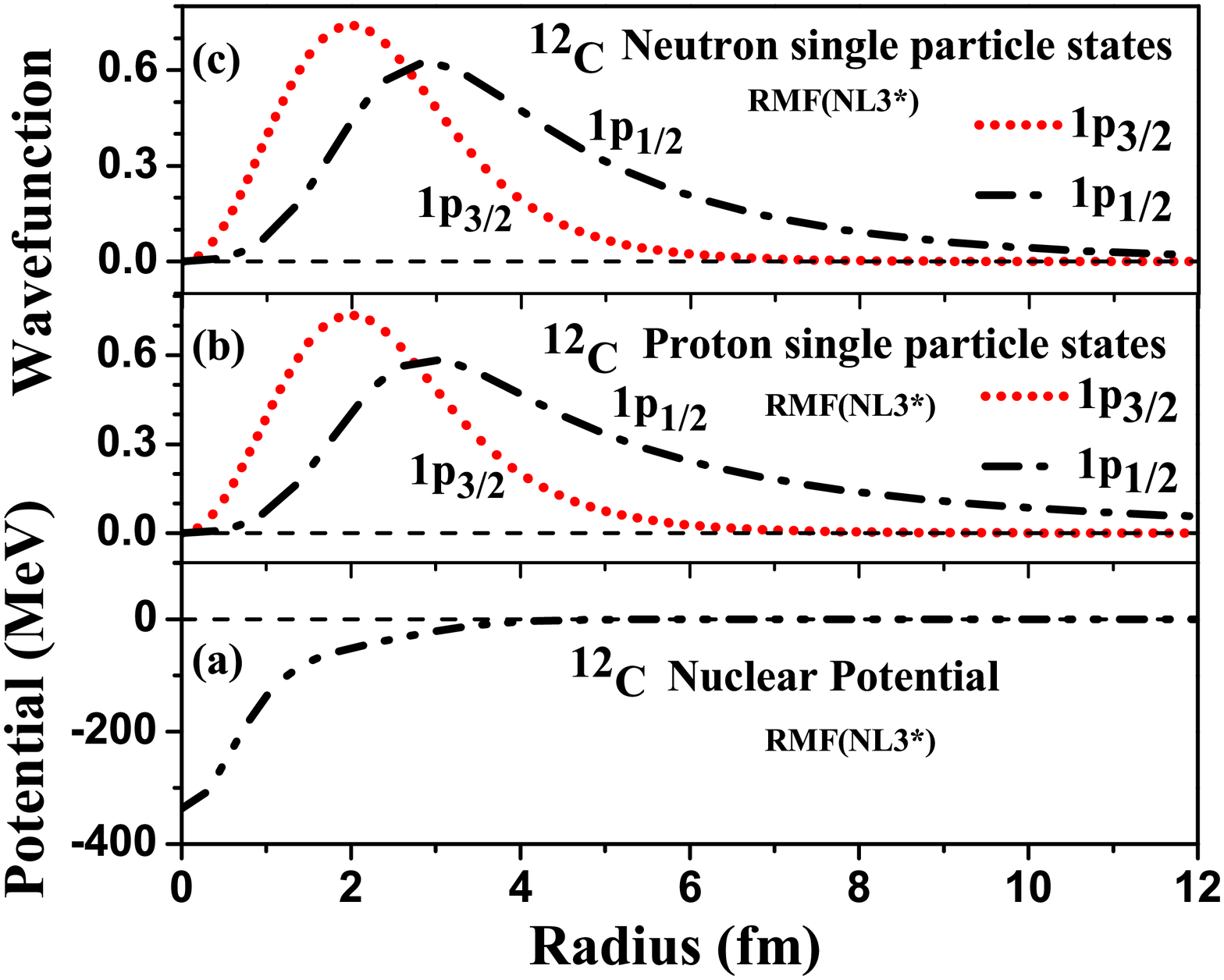}
\caption{(Colour online) Sum of scalar and vector potentials (a),  wave function of proton (b) and neutron (c) single particle 1p$_{3/2}$ and 1p$_{1/2}$ states.} \label{fig6} \end{figure}

Further, to assure magicity of Z = 6, in Fig. \ref{fig6}  we have plotted RMF potential, a sum of scalar and vector potentials, for $^{12}$C in Fig. \ref{fig6}(a) and wavefunctions for $^{12}$C of proton and neutron single particle states 1p$_{3/2}$ and 1p$_{1/2}$ in Figs. \ref{fig6}(b) and (c) respectively. The very large energy gap between 1p$_{3/2}$ and 1p$_{1/2}$ as obtained in Fig. \ref{fig5}, can be testified with the different behaviour of wave functions for the 1p$_{3/2}$ and 1p$_{1/2}$ states. In Fig. \ref{fig6}(b) and (c) the wavefunction of 1p$_{3/2}$ is clearly seen to be confined within a radial range of about 4-5 fm (Fig. \ref{fig6}(a)). In contrast, wavefunction of 1p$_{1/2}$ is seen to be spread over outside the potential region and hence providing no chance for the particles to occupy 1p$_{1/2}$ state.

\begin{figure}[h]
\centering
\includegraphics[width=0.5\textwidth]{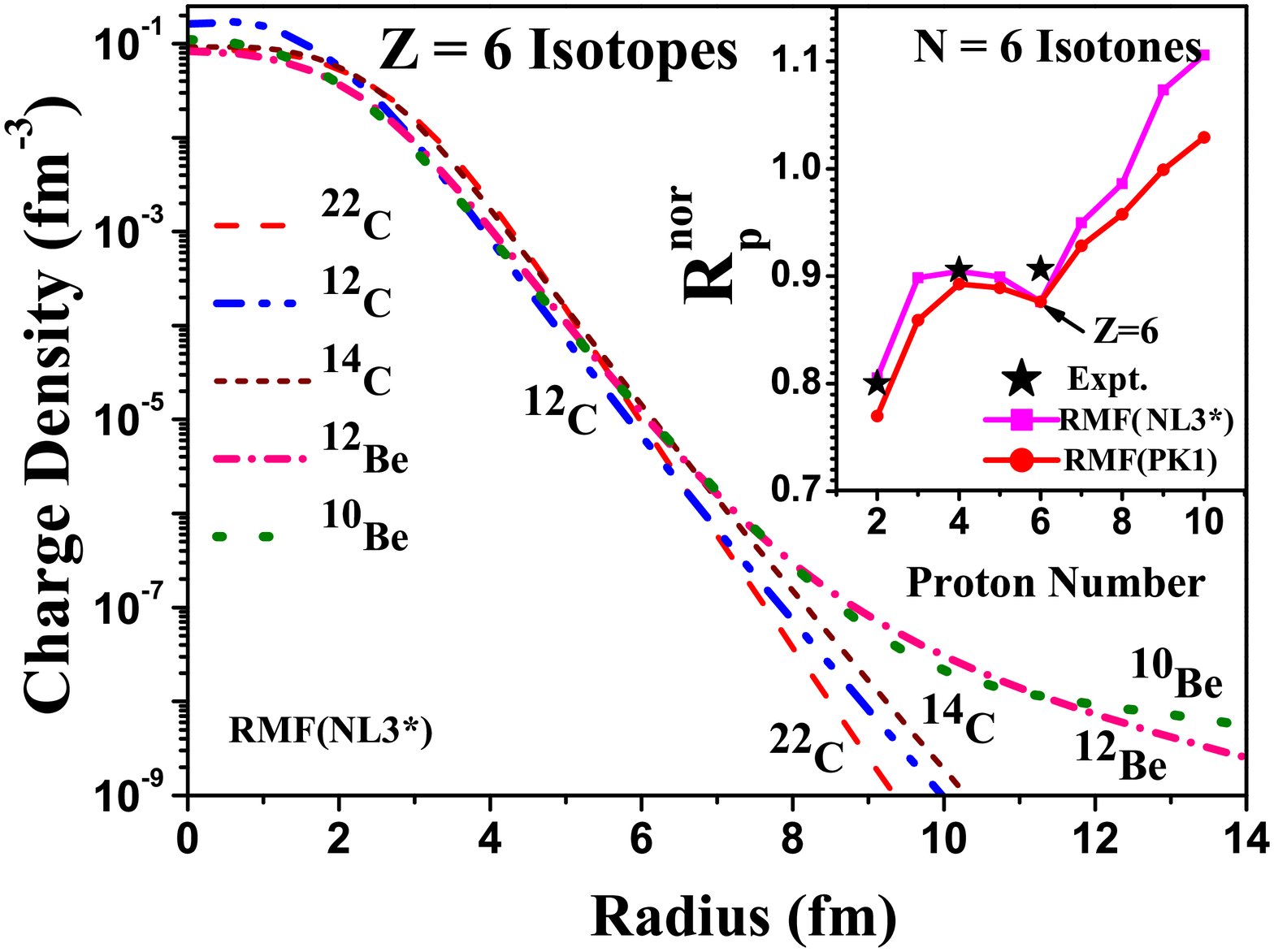}
\caption{(Colour online) Radial distribution of charge density of $^{12,14,22}$C are shown along with charge density of $^{10,12}$Be. Inset shows normalized radii (R$_{p}^{nor}$ = R$_{p}$/R$_{p}^{Co}$) for N = 6 isotones calculated by NL3* and PK1 parameters. Experimental data are taken from \cite{Tanihata,Liatarbz,Kanungo1}.} \label{fig7} \end{figure}

 We demonstrate radial distribution of charge density of $^{12,14,22}$C along with $^{10,12}$Be in Fig. \ref{fig7}. It is indulging to see from Fig. \ref{fig7} that charge densities of nuclei with Z = 6 show sharp fall and confined to smaller distances compared to those for the charge density distributions of nuclei with lower Z ($^{10,12}$Be). This sharp fall in asymptotic density values is due to the fact that for the closed shell isotopes (here Z = 6) there are no contributions to the density from the quasi bound states having positive energy near the continuum threshold. Similar pattern of density (neutron density) we have found for isotones of N = 6 (not shown here) and these distributions are in support of magicity of Z = 6 and N = 6.

In addition, we have plotted proton rms radii for N = 6 isotones calculated by NL3* and PK1 parameters. To eliminate the smooth mass number dependence of the proton rms radii R$_{p}$, we normalized proton rms radii using the following formula given by Collard \textit{et al.} \cite{collard}

\begin{equation}
R_{p}^{Co} = \sqrt{3/5}\,\, (1.15 + 1.80A^{-2/3} - 1.20A^{-4/3})\,\,A^{1/3}
\end{equation}

Therefore, the normalized radii (R$_{p}^{nor}$ = R$_{p}$/R$_{p}^{Co}$) is plotted in the inset of Fig. \ref{fig7}. Angeli \textit{et al.} have suggested a kink (change in the slope) particularly for shell closure \cite{angeli,angeli2} which is evident here also in inset of Fig. \ref{fig7} for the case of Z = 6. This kind of kink observed for Z = 6 is another demonstration of its magicity. Various ground state properties obtained with NL3* and PK1 parameters are mentioned in Table II with available experimental data \cite{angeli,neumaier,Tanihata,Liatarbz,Ozawa,Kanungo1} depicting comprehensive study and general validity of RMF+BCS approach in this light mass region.

\begin{table*}
\caption{Binding energy, charge radius $R_{c}$, neutron radius $R_{n}$, proton radius $R_{p}$, and matter radius $R_{m}$ for the nuclei with Z(N) = 6 are tabulated and compared with available experimental data which are taken from Refs.$~$ \cite{angeli,neumaier,Tanihata,Liatarbz,Ozawa,Kanungo1}.}
\centering
\textbf{
\resizebox{1.0\textwidth}{!}{%
{\begin{tabular}{c|ccc|ccc|ccc|ccc|ccc}
 \hline
 \multicolumn{1}{c|}{Nuclei}&
 \multicolumn{3}{c|}{Binding Energy (MeV)}&
 \multicolumn{3}{c|}{$R_{c}$ (fm) }&
 \multicolumn{3}{c|}{$R_{n}$ (fm)}&
 \multicolumn{3}{c|}{$R_{p}$ (fm) }&
 \multicolumn{3}{c}{$R_{m}$ (fm) }\\
\cline{2-16}
 \multicolumn{1}{c|}{}&
 \multicolumn{1}{c}{NL3*}&
 \multicolumn{1}{c}{PK1}&
  \multicolumn{1}{c|}{Expt.}&
   \multicolumn{1}{c}{NL3*}&
    \multicolumn{1}{c}{PK1}&
    \multicolumn{1}{c|}{Expt.}&
       \multicolumn{1}{c}{NL3*}&
       \multicolumn{1}{c}{PK1}&
       \multicolumn{1}{c|}{Expt.}&
       \multicolumn{1}{c}{NL3*}&
 \multicolumn{1}{c}{PK1}&
  \multicolumn{1}{c|}{Expt.}&
  \multicolumn{1}{c}{NL3*}&
 \multicolumn{1}{c}{PK1}&
  \multicolumn{1}{c}{Expt.}\\
   \hline
  $^{11}${C}&73.05   &73.44   &73.44 &2.52 &2.48&    & 2.26 & 2.23 &    &2.39&2.35&    &2.34&2.30 &  \\
  $^{12}${C}&89.86   &90.00   &92.16 &2.42&2.39&2.47$\pm$0.002& 2.25& 2.23 &2.49$\pm$0.01&2.28&2.25&2.32$\pm$0.02&2.28&2.24&2.35$\pm$0.02\\
  $^{13}${C}&97.30   &97.42   &97.11 &2.47 &2.44&2.46$\pm$0.003& 2.45 & 2.40 &    &2.33&2.31&2.30$\pm$0.04&2.40&2.36 &2.28$\pm$0.04\\
  $^{14}${C}&106.11  &106.21  &105.28&2.52 &2.48&2.50$\pm$0.009& 2.58 & 2.52 &2.70$\pm$0.10&2.39&2.35&2.32$\pm$0.04&2.50&2.45 &2.33$\pm$0.07\\
  $^{15}${C}&107.67  &107.07  &106.50&2.54 &2.51&    & 2.79 & 2.66 &    &2.41&2.38&2.37$\pm$0.03&2.64&2.55 &2.54$\pm$0.04\\
  $^{16}${C}&110.99  &110.37  &110.75&2.56 &2.53&    & 2.90 & 2.77 &2.88$\pm$0.09&2.43&2.40&2.40$\pm$0.04&2.73&2.64 &2.74$\pm$0.03\\
  $^{17}${C}&113.07  &111.96  &111.49&2.59 &2.56&    & 3.01 & 2.86 &    &2.46&2.43&2.42$\pm$0.04&2.83&2.72 &2.76$\pm$0.03\\
  $^{18}${C}&116.16  &114.83  &115.67&2.61 &2.59&    & 3.12 & 2.96 &3.06$\pm$0.29&2.49&2.46&2.39$\pm$0.04&2.92&2.80 &2.86$\pm$0.04\\
  $^{19}${C}&118.40  &116.12  &116.24&2.63 &2.61&    & 3.23 & 3.04 &    &2.51&2.48&2.40$\pm$0.03&3.02&2.88 &3.16$\pm$0.07\\
  $^{20}${C}&121.40  &119.33  &119.18&2.67 &2.65&    & 3.26 & 3.11 &    &2.54&2.52&    &3.06&2.94 &2.98$\pm$0.05\\
  $^{21}${C}&121.71  &118.35  &119.15&2.65 &2.64&    & 3.37 & 3.17 &    &2.53&2.51&    &3.15&3.00 &    \\
  $^{22}${C}&122.95  &118.78  &119.28&2.58 &2.57&    & 3.35 & 3.20 &    &2.45&2.44&    &3.13&3.01 &    \\
  \hline
  $^{8}${He}&  37.45 &34.96  & 31.40 & 2.06&1.99 &1.92$\pm$0.03&2.84& 2.51 &2.50$\pm$0.19&1.90 &1.82&1.89$\pm$0.17&2.63&2.36 &2.37$\pm$0.18\\
  $^{9}${Li}&  49.56 &48.60  & 45.34 & 2.32&2.23 &2.25$\pm$0.05&2.65& 2.45 &    &2.17 &2.08&    &2.50&2.33 &    \\
  $^{10}${Be}& 64.21 &64.31  & 64.98 & 2.38&2.35 &2.36$\pm$0.02&2.47& 2.40 &2.58$\pm$0.04&2.24 &2.21&2.24$\pm$0.02&2.38&2.33 &    \\
  $^{11}${B}&  75.95 &76.39  & 76.20 & 2.41&2.38 &2.41$\pm$0.03&2.35& 2.32 &    &2.27 &2.24&    &2.32&2.29 &    \\
  $^{12}${C}&89.86   &90.00   &92.16 &2.42&2.39&2.47$\pm$0.002& 2.25& 2.23 &2.49$\pm$0.01&2.28&2.25&2.32$\pm$0.02&2.28&2.24&2.35$\pm$0.02\\
  $^{13}${N}&  94.17 &94.17  & 94.11 & 2.61&2.56 &    &2.31& 2.29 &    &2.48 &2.43&    &2.41&2.37 &    \\
  $^{14}${O}&  99.42 &99.32  & 98.73 & 2.74&2.67 &    &2.38& 2.34 &    &2.62 &2.55&    &2.52&2.46 &    \\
  $^{15}${F}&  97.71 &96.52  & 97.22 & 3.01&2.81 &    &2.39& 2.37 &    &2.90 &2.70&    &2.71&2.57 &    \\
  $^{16}${Ne}& 97.16 &95.68  & 97.33 & 3.13&2.93 &    &2.43& 2.40 &    &3.03 &2.82&    &2.82&2.67 &    \\
\hline
\end{tabular}}}}
\end{table*}

\section{Conclusions}

A systematic study using relativistic mean-field plus BCS approach to investigate magicity in Z = 6 isotopes and N = 6 isotones has been done with the help of ground state properties. These calculations have been performed first assuming an axially deformed shape (deformed RMF) and also for zero deformation by employing spherical RMF approach which include binding energies, deformations, proton and neutron pairing energies, two neutron and proton separation energies, single particle states as well as radii and density profiles calculated by mainly NL3* parameter \cite{nl3star} and compared by PK1 \cite{pk1} parameter. Results of both these parameters are found in excellent match with each other and available experimental data.

A detailed analysis of binding energies, deformations, pairing energies and one- and two-proton(neutron) separation energies establishes $^{12,14,22}$C, $^8$He and $^{14}$O as doubly magic candidates with either Z = 6 or N =6. Magic character of Z = 6 protons and N = 6 neutrons is observed consistently which is verified by a large gap between 1p$_{3/2}$ and 1p$_{1/2}$ single particle levels and dissimilar behaviour of wavefunctions of these states. Charge density distribution of C isotopes is also found to confined to a smaller distance comparative to charge distribution of nuclei with smaller Z and a kink in proton radii demonstrates proton magicity in Z = 6 in accord with recent observation by Tran {\textit{et al}.}. In a similar way, magicity of N = 6 is also reported with the same evidences which may be a testing ground for upcoming experiments.

\section{Acknowledgement}
Authors would like to thank Prof. H. L. Yadav, Banaras Hindu University,
Varanasi, India and Prof. L. S. Geng, Beihang University, China
for their guidance. One of the authors (G. Saxena)
gratefully acknowledges the support provided by Science and Engineering Research Board (DST),
India under the young scientist project YSS/2015/000952.

\end{document}